\documentclass[preprint,aps,prd,showpacs,superscriptaddress,nofootinbib]{revtex4}

\usepackage{amsfonts}
\usepackage{mathrsfs}
\usepackage{graphicx}
\usepackage{amsmath}
\usepackage{amssymb}
\usepackage{bm}
\usepackage{bbm}

\def\bfalpha{\mbox{\boldmath $\alpha$}}

\def\bfsigma{\mbox{\boldmath $\sigma$}}

\def\bfvarepsilon{\mbox{\boldmath $\varepsilon$}}

\def\OMIT#1{}

\newcommand{\nn}{\nonumber}

\newcommand{\beq}{\begin{equation}}
\newcommand{\eeq}{\end{equation}}
\newcommand{\bqa}{\begin{eqnarray}}
\newcommand{\eqa}{\end{eqnarray}}

\begin{document}

\title{\mbox{}\\[10pt]
$\bm{\mathcal O}(\bfalpha_{\bm s} \bm{v^2})$ correction to pseudoscalar quarkonium decay to two photons
}


\author{Yu Jia\footnote{E-mail: jiay@ihep.ac.cn}}
\affiliation{Institute of High Energy Physics, Chinese Academy of
Sciences, Beijing 100049, China\vspace{0.2cm}}
\affiliation{Theoretical Physics Center for Science Facilities, Chinese Academy of
Sciences, Beijing 100049, China\vspace{0.2cm}}

\author{Xiu-Ting Yang\footnote{E-mail: yangxt@ihep.ac.cn}}
\affiliation{Institute of High Energy Physics, Chinese Academy of
Sciences, Beijing 100049, China\vspace{0.2cm}}

\author{Wen-Long Sang\footnote{E-mail: swlong@korea.ac.kr}}
\affiliation{Theoretical Physics Center for Science Facilities,
Chinese Academy of Sciences, Beijing 100049, China\vspace{0.2cm}}
\affiliation{Department of Physics, Korea University, Seoul 136-701,
 Korea\vspace{0.2cm}}

\author{Jia Xu\footnote{E-mail: xuj@ihep.ac.cn}}
\affiliation{Institute of High Energy Physics, Chinese Academy of
Sciences, Beijing 100049, China\vspace{0.2cm}}

\date{\today}
\begin{abstract}

We investigate the ${\cal O}(\alpha_s v^2)$ correction to the process of
pseudoscalar quarkonium decay to two photons
in nonrelativistic QCD (NRQCD) factorization framework.
The short-distance coefficient associated with the relative-order $v^2$ NRQCD matrix element
is determined to next-to-leading order in $\alpha_s$ through the perturbative matching procedure.
Some technical subtleties encountered in calculating the ${\cal O}(\alpha_s)$ QCD amplitude
are thoroughly addressed.

\end{abstract}



\maketitle

\section{Introduction}

Quarkonium inclusive annihilation decays are historically among the
earliest applications of perturbative quantum chromodynamics
(QCD)~\cite{Appelquist:1974zd,DeRujula:1974nx,Barbieri:1975am,Novikov:1977dq}.
At present, it is widely accepted that these quarkonium inclusive
decay processes can be systematically described by the
nonrelativistic QCD (NRQCD) factorization
approach~\cite{Bodwin:1994jh}, which is based on the
effective-field-theory formalism and directly linked with the first
principles of QCD.

In NRQCD factorization approach, the inclusive decay rate can be
systematically expressed as the sum of product of short-distance
coefficients and the NRQCD operator matrix elements. The
short-distance coefficients encode the hard effects of quark and
antiquark annihilation at length scale of order $1/m$ ($m$ denotes
the mass of heavy quark), which therefore can be accessed by
perturbation theory owing to asymptotic freedom of QCD. In contrast,
the NRQCD matrix elements are sensitive to the nonperturbative
dynamics that occurs at distance of $1/mv$ or longer ($v$ denotes
the typical (anti-)quark velocity inside a quarkonium). An important
feature of these nonperturbative matrix elements is that they are
universal, and satisfy a definite power-counting in $v$. This
attractive feature endows the NRQCD factorization approach with a
controlled predictive power.

Among the quarkonium annihilation decay processes, the simplest and
cleanest are the quarkonium electromagnetic decays, exemplified by
vector quarkonium decay to a lepton pair and pseudoscalar quarkonium
decay to two photons. Both of these two processes have been
comprehensively studied in theory and experiment for decades. In
this work, we will address the ${\cal O}(\alpha_s v^2)$ correction
to the latter process in NRQCD factorization context. We note that
investigations on this process from other approaches are also
available (e.g.,
see~\cite{Ebert:2003mu,Lansberg:2006dw,Yang:2009kq}).

We will generically label a pseudoscalar quarkonium by $\eta_Q$,
where $Q$ can stand for the $c$ or the $b$ quark. Through relative
order-$v^4$, the NRQCD factorization formula for the decay rate of
$\eta_Q\to\gamma\gamma$ reads~\cite{Bodwin:2002hg,Brambilla:2006ph}:
\bqa
\Gamma[\eta_Q\to \gamma\gamma]& = & {F({}^1S_0)\over m^2}
\left|\langle 0 |\chi^\dagger\psi |\eta_Q\rangle\right|^2 +
{G({}^1S_0)\over m^4}{\rm Re}\bigg\{\langle \eta_Q|\psi^\dagger\chi|0\rangle \langle 0| \chi^\dagger
(-\tfrac{i}{2}\tensor{\mathbf{D}})^{2}\psi| \eta_Q\rangle \bigg\}
\nn\\
&+& {H^1({}^1S_0)\over m^6} \langle \eta_Q|\psi^\dagger
(-\tfrac{i}{2}\tensor{\mathbf{D}})^{2} \chi|0\rangle
\langle 0| \chi^\dagger
(-\tfrac{i}{2}\tensor{\mathbf{D}})^{2}\psi| \eta_Q\rangle
\nn\\
&+&
 {H^2({}^1S_0)\over m^6}{\rm Re}
 \bigg\{\langle \eta_Q|\psi^\dagger \chi|0\rangle \langle 0|
 \chi^\dagger (-\tfrac{i}{2}\tensor{\mathbf{D}})^{4} \psi| \eta_Q\rangle
 \bigg\}+O(v^6\Gamma),
\label{NRQCD:factorization:etaQ:decay:width}%
\eqa
where $\psi$ and $\chi^\dagger$ represent Pauli spinor fields that
annihilate the heavy quark $Q$ and heavy antiquark $\overline{Q}$,
respectively.

The short-distance coefficients $F$, $G$, $H$ in
(\ref{NRQCD:factorization:etaQ:decay:width})
are
\begin{subequations}
\bqa
& & F({}^1S_0) =  2\pi e_Q^4 \alpha^2\left[1+C_F{\alpha_s\over\pi}\left({\pi^2\over 4}-5\right)+{\cal O}(\alpha_s^2)\right],
\\
& & G({}^1S_0) = -{8\pi e_Q^4 \alpha^2\over 3}\bigg[1+{\cal O}(\alpha_s)\bigg],
\\
& & H^1({}^1S_0)+H^2({}^1S_0) = {136 \pi e_Q^4 \alpha^2\over
45}\bigg[1+{\cal O}(\alpha_s)\bigg],
\label{coeff:H1+H2:tree:level}
\eqa
\end{subequations}
where $e e_Q$ denotes the electric charge of the heavy quark $Q$.
The ${\cal O}(\alpha_s)$ correction to $F({}^1S_0)$ was first
computed in \cite{Harris:1957zz,Barbieri:1979be,Hagiwara:1980nv}. An
incomplete ${\cal O}(\alpha_s^2)$ correction to this coefficient has
also been available about a decade ago~\cite{Czarnecki:2001zc},
which indicates an uncomfortably large negative correction. The
coefficient $G({}^1S_0)$ is associated with the order-$v^2$ matrix
element, whose tree-level value has been known long
ago~\cite{Keung:1982jb}. Recently, the tree-level coefficients $H({}^1S_0)$
associated with the order-$v^4$ matrix elements are also available
for the first time~\cite{Bodwin:2002hg,Brambilla:2006ph}.
Initially only the combination of two ${\cal O}(v^4)$ coefficients was given~\cite{Bodwin:2002hg},
as shown in (\ref{coeff:H1+H2:tree:level}).
Later Ref.~\cite{Brambilla:2006ph} was able to determine
these two coefficients separately: $ H^1({}^1S_0 )={20 \pi e_Q^4
\alpha^2\over 9}$ and $H^2({}^1S_0)={4 \pi e_Q^4 \alpha^2\over 5}$.

In recent years there have also been many phenomenological
investigations on the nonperturbative NRQCD matrix elements for
the pseudoscalar quarkonium state~\cite{He:2007te,Bodwin:2007fz,Chung:2010vz}.
Most of them focus on the matrix elements at lowest order (LO) and next-to-leading order (NLO)
in $v^2$.
Comparing equation (\ref{NRQCD:factorization:etaQ:decay:width}) truncated at order $v^2$
with the measured decay rate of $\eta_c\to\gamma\gamma$, some authors are able to
fit the first two NRQCD matrix elements of $\eta_c$ and find they roughly obey the velocity
counting rules~\cite{He:2007te,Bodwin:2007fz}~\footnote{The measured partial width of $\eta_c\to\gamma\gamma$
seems to have changed significantly in past few years, though the full width of $\eta_c$
almost remains intact.
For example, the branching fraction of this decay channel was reported to be
$(2.4_{-0.9}^{+1.1})\times 10^{-4}$
in PDG 2008 edition~\cite{Amsler:2008zzb}. But this value has reduced to
$(6.3\pm 2.9)\times 10^{-5}$ in the latest PDG 2010 edition~\cite{Nakamura:2010zzi}.
This may cast some shadow on
the reliability of the fitted values for $\eta_c$ NRQCD matrix elements
by using older data~\cite{He:2007te,Bodwin:2007fz}.}.

By far, the ${\cal O}(\alpha_s^2 v^0)$
contribution~\cite{Czarnecki:2001zc} and ${\cal O}(\alpha_s^0 v^4)$
contribution~\cite{Bodwin:2002hg,Brambilla:2006ph} to the decay rate
of $\eta_Q\to \gamma\gamma$ are known. Assuming $\alpha_s(m)\sim
v^2$, one may naturally wonder what is the actual size of the ${\cal
O}(\alpha_s v^2)$ correction. In order to answer this question, one
needs first know the ${\cal O}(\alpha_s)$ correction to the
short-distance coefficient $G({}^1S_0)$. It is the purpose of this
work to compute this correction through the perturbative matching
method.

The remainder of this paper is organized as follows. In
section~\ref{matching:stragey}, we outline the perturbative matching
strategy which can be utilized to deduce the NRQCD short-distance coefficients for our process.
In section~\ref{sec:tech:full:QCD}, we elaborate on some technical
issues encountered in calculating the ${\cal O}(\alpha_s)$
correction to $Q\overline{Q}({}^1S_0)\to\gamma\gamma$ with the
covariant projection approach. In particular, we specify the
prescription of $\gamma_5$ in dimensional regularization adopted in
this work. We also mention some technical ambiguities about
extracting the $S$-wave amplitude.
In section~\ref{sec:full:QCD:NLO}, we then employ the covariant
projection technique to compute the $Q\overline{Q}({}^1S_0)\to
\gamma\gamma$ amplitude to NLO in $\alpha_s$.
The corresponding ${\cal O}(\alpha_s)$ calculation in the NRQCD side
is presented in section~\ref{sec:NRQCD:NLO}.
In section~\ref{Matching:short-distance:NLO}, by comparing the NLO
QCD amplitude and the respective NRQCD amplitude, we determine the
first two short-distance coefficients in
(\ref{NRQCD:factorization:etaQ:decay:width}) through order
$\alpha_s$.
Finally in section~\ref{summary}, we present a brief summary. The
appendix is devoted to enumerating the analytic expressions of the
one-loop scalar integrals encountered in the calculation of the NLO
QCD correction to $Q\overline{Q}({}^1S_0)\to\gamma\gamma$.

\section{The method of deducing the short-distance coefficients
\label{matching:stragey}}

In many effective field theories, it is a standard procedure to
determine the short-distance coefficient through the matching
procedure. We will follow this orthodox method in this work. Our
calculation is in close analogy with Ref.~\cite{Luke:1997ys}, where
the ${\cal O}(\alpha_s v^2)$ correction for $J/\psi\to e^+e^-$ has
been deduced also using the matching approach.

It is worth mentioning that the alternative, even more efficient way of deducing
the short-distance coefficients exists, exemplified by the {\it method of region}
developed by Beneke and Smirnov~\cite{Beneke:1997zp}.
This approach allows one to dissect the loop integral for
quarkonium annihilation process into
four distinct regions: {\it hard}, {\it soft}, {\it potential} and {\it ultrasoft}.
The short-distance coefficient
only receives the contribution from the hard region,
while the NRQCD effective theory
characterizes the dynamics from the three low-energy regions.
To determine the short-distance coefficient,
one can either directly compute the hard region contribution order by order in $v$ expansion,
or by first calculating the full QCD diagram, then subtracting the contributions from three low
energy regions. The second strategy turns out to have some great technical advantage, which enables one
to deduce a class of relativistic corrections to arbitrary orders in $v^2$ with ease.
Following this spirit, there have recently been
progresses in inferring the all-order-in-$v^2$ corrections to
$J/\psi\to e^+e^-$~\cite{Bodwin:2008vp} and $B_c\to e \bar\nu_e$\cite{Lee:2010ts}.

\subsection{NRQCD factorization formula for the decay amplitude and width
\label{matching:stragey:mine}}

Let us label the momenta of $\eta_Q$ and two photons by $P$, $k_1$ and $k_2$,
and the polarization vectors of two photons by $\varepsilon_1$ and $\varepsilon_2$, respectively.
The Lorentz and parity invariance dictates the amplitude of $\eta_Q \to \gamma\gamma$
uniquely of the structure $\epsilon_{\alpha\beta\mu\nu} P^\alpha k_1^\beta
\varepsilon_1^{*\mu} \varepsilon_2^{*\nu}$.
For the problem at hand, it is most natural to work in the $\eta_Q$ rest frame.
In this frame, the amplitude is then proportional to the kinematic invariant
${\mathbf k}_1 \cdot \bfvarepsilon_1^* \times \bfvarepsilon_2^*$.

When considering quarkonium electromagnetic annihilation decay,
one can directly invoke NRQCD factorization at the amplitude level.
We need only retain those NRQCD color-singlet operator matrix elements that connect the
vacuum to the $\eta_Q$ state. To the order of desired
accuracy, one expects that the following factorization formula holds:
\bqa
{\mathcal A}[\eta_Q \to \gamma\gamma] &=&
\hat{\mathbf k}_1 \cdot \bfvarepsilon_1^* \times \bfvarepsilon_2^*
\left[ c_0 \langle 0| \chi^\dagger \psi|\eta_Q \rangle
 + {c_2 \over m^2} \langle 0| \chi^\dagger
(-\tfrac{i}{2}\tensor{\mathbf{D}})^{2}\psi| \eta_Q\rangle + {\cal O}(v^4)
\right],
\label{NRQCD:factorization:ampl:level}%
\eqa
where $c_i$ ($i=0,2$) signify the short-distance coefficients
associated with the NRQCD matrix elements at LO and NLO in $v^2$. It
is especially convenient in this work to adopt the nonrelativistic
normalization for $\eta_Q$ state in both sides of
(\ref{NRQCD:factorization:ampl:level}). In addition, we have
explicitly factored out the kinematic invariant, which is
represented by a dimensionless Lorentz pseudoscalar,
\bqa
\hat{\mathbf k}_1 \cdot \bfvarepsilon_1^* \times \bfvarepsilon_2^* &
= & -{2\over M_{\eta_Q}^2} \epsilon_{\alpha\beta\mu\nu} P^\alpha
k_1^\beta \varepsilon_1^{*\mu} \varepsilon_2^{*\nu},
\label{cross:product:Lorentz:structure}%
\eqa
in the right-hand side of (\ref{NRQCD:factorization:ampl:level}),
where $\hat{\mathbf k}_1={\mathbf k}_1/|{\mathbf k}_1|$ is a unit
vector. The separation of this kinematic factor renders $c_i$
($i=0,2$) to bear a very simple form.

Squaring the amplitude in (\ref{NRQCD:factorization:ampl:level}),
summing over the polarizations of photons, and
integrating over the phase space and accounting for the indistinguishability of two photons,
we can express the decay rate of $\eta_Q \to \gamma\gamma$ as
\bqa
\Gamma[\eta_Q \to \gamma\gamma] &=&
{1\over 2!} \int \!\!
{d^{3}k_1 \over (2\pi)^{3}2k_1^0} {d^{3}k_2 \over (2\pi)^{3}2k_2^0}
(2\pi)^{4} \delta^{(4)}(P-k_1-k_2) \sum \left|\mathcal{A}[\eta_Q \to \gamma\gamma]\right|^2
\nn\\
&=& {1\over 8\pi}
\left| c_0 \langle 0| \chi^\dagger \psi|\eta_Q \rangle
 + {c_2 \over m^2} \langle 0| \chi^\dagger
(-\tfrac{i}{2}\tensor{\mathbf{D}})^{2}\psi| \eta_Q\rangle + \cdots
\right|^2,
\label{NRQCD:factorization:decay:rate:level}%
\eqa
where the formula $\sum_{\rm Pol} \, \left|\hat{\mathbf k}_1 \cdot \bfvarepsilon_1^* \times \bfvarepsilon_2^*
\right|^2 = 2$ has been used.

Comparing (\ref{NRQCD:factorization:decay:rate:level}) with (\ref{NRQCD:factorization:etaQ:decay:width}),
one can express the short-distance coefficients $F({}^1S_0)$ and $G({}^1S_0)$ that appear in (\ref{NRQCD:factorization:etaQ:decay:width}), the standard NRQCD
factorization formula for the decay rate, in terms of $c_0$ and $c_2$:
\begin{subequations}
\bqa
F({}^1S_0)& = &  {m^2 \over 8\pi} |c_0|^2,
\\
G({}^1S_0)& = &  {m^2\over 4\pi} {\rm Re}[c_0 c_2^*].
\eqa
\label{Relation:F:G:c0:c2}
\end{subequations}
The goal of this work is to determine $G({}^1S_0)$ through the order $\alpha_s$.
Therefore, we first need determine both of the coefficients $c_0$ and $c_2$ to order $\alpha_s$.

\subsection{Matching at the amplitude level
\label{matching:stragey:mine}}

To determine the values of $c_0$ and $c_2$, we follow the moral that
these short-distance coefficients, encapsulating the {\it hard}
quantum fluctuations that occur at the length scale $\sim 1/m$, are
insensitive to the long-distance hadronic dynamics. As a convenient
calculational device, one can replace the physical $\eta_Q$ meson by
a free $Q\overline{Q}$ pair of the quantum number ${}^1S_0^{[1]}$,
so that both the full amplitude, ${\mathcal
A}[Q\overline{Q}({}^1S_0^{[1]})\to\gamma\gamma]$, and the NRQCD
operator matrix elements can be accessed by perturbation theory. The
short-distance coefficients $c_i$ then can be solved by equating the
QCD amplitude ${\mathcal A}$ and the corresponding NRQCD amplitude
${\mathbb A}_{\rm NRQCD}$, order by order in $\alpha_s$. This
procedure is commonly referred to as {\it perturbative matching}.
Analogous to (\ref{NRQCD:factorization:ampl:level}), one can write
down the pertubative matching formula for the
$Q\overline{Q}({}^1S_0^{[1]})\to\gamma\gamma$ process:
\begin{subequations}
\bqa
& & {\mathcal A}[Q\overline{Q}({}^1S_0^{[1]}) \to \gamma\gamma] =
\hat{\mathbf k}_1 \cdot \bfvarepsilon_1^* \times \bfvarepsilon_2^*
\, {\mathbb A}_{\rm NRQCD}, \label{Matching:QQbar:2gamma:ampl:level}
\\
& & {\mathbb A}_{\rm NRQCD} =  c_0 \langle 0| \chi^\dagger
\psi|Q\overline{Q}({}^1S_0^{[1]})\rangle
 + {c_2 \over m^2} \langle 0| \chi^\dagger
(-\tfrac{i}{2}\tensor{\mathbf{D}})^{2}\psi|Q\overline{Q}({}^1S_0^{[1]})\rangle
+ \cdots. \label{QQbar:2gamma:ampl:level:NRQCD:ampl}
\eqa
\label{Matching:formula:ampl:QQbar:2gamma}
\end{subequations}
In Eq.~(\ref{Matching:formula:ampl:QQbar:2gamma}), we again adopt the nonrelativistic normalization for the $Q$
and $\overline{Q}$ states in the computations of the full QCD amplitude
and the NRQCD matrix elements.

One can organize the full amplitude $\mathcal{A}$ in powers of the relative momentum between $Q$ and $\overline{Q}$,
denoted by ${\bf q}$:
\bqa
\mathcal{A}[Q\overline{Q}({}^1S_0^{[1]})\to \gamma\gamma] =
\hat{\mathbf k}_1 \cdot \bfvarepsilon_1^* \times \bfvarepsilon_2^*
\left[{\mathscr A}_0+ {{\bf q}^2 \over m^2}{\mathscr A}_2+ {\cal
O}(\bf{q}^4) \right]. \label{full:amplitude:S:wave:series}
\eqa
For convenience, we have again factored out the kinematic invariant
$\hat{\mathbf k}_1 \cdot \bfvarepsilon_1^* \times \bfvarepsilon_2^*$ in the amplitude.
To the desired accuracy, one can truncate the series at ${\cal O}({\bf q}^2)$,
with the first two Taylor coefficients denoted by ${\mathscr A}_0$ and ${\mathscr A}_2$.
To our purpose, we need compute both ${\mathscr A}_i$ through NLO in $\alpha_s$.
This will be conducted in Section~\ref{sec:full:QCD:NLO}.

The NRQCD amplitude ${\mathbb A}_{\rm NRQCD}$ in
(\ref{Matching:formula:ampl:QQbar:2gamma}), or equivalently, the
NRQCD vacuum-to-$Q\overline{Q}$ matrix elements, need also be worked
out in perturbation theory through ${\cal O}(\alpha_s)$.  The
encountered NRQCD matrix elements at LO in $\alpha_s$ are
particularly simple~\footnote{Unless otherwise stated, throughout
this work we use the superscripts $(0)$ and $(1)$ to indicate the LO
and NLO contributions in $\alpha_s$, and the subscripts $0$ and $2$
to represent the LO and NLO contributions in $v^2$.}:
\begin{subequations}
\label{NRQCD:matrix:elements:LO}
\bqa
& & \langle0\vert\chi^\dagger\psi\vert
Q\overline{Q}({}^1S_0^{[1]})\rangle^{(0)} = \sqrt{2N_c},\\
& &
\langle0\vert\chi^\dagger(-\frac{i}{2}\tensor{\bold{D}})^2\psi\vert
Q\overline{Q}({}^1S_0^{[1]}) \rangle^{(0)} = \sqrt{2 N_c}\,{\bf
q}^2,
\eqa
\end{subequations}
where the factor $\sqrt{2N_c}$ is due to the spin and color factors
of the normalized $Q\overline{Q}({}^1S_0^{[1]})$ state. The
computation of these matrix elements to ${\cal O}(\alpha_s)$ will be
addressed in Section~\ref{sec:NRQCD:NLO}.

\section{Techniques about computing full QCD amplitude}
\label{sec:tech:full:QCD}

In this section, we outline some necessary techniques about
calculating the amplitude for $Q\overline{Q}(^1S_0^{[1]})\to
\gamma\gamma$.

\subsection{Kinematic setup}
\label{kinematics:definition}

Let $p$ and $\bar{p}$ represent the momenta carried by $Q$ and $\overline{Q}$, respectively.
It is customary to decompose them in the following form:
\begin{subequations}
\bqa
p &=& {1 \over 2}P+q,%
\\
\bar{p} &=& {1 \over 2}P-q,
\eqa
\label{p-pbar-mom}%
\end{subequations}
where $P$ is the total momentum of the $Q\overline{Q}(^1S_0^{[1]})$
pair with invariant mass $\sqrt{P^2}=2 E$, $q$ is the relative
momentum. Enforcing $Q$ and $\overline{Q}$ to stay on their mass
shells, one requires that $E=\sqrt{m^2-q^2}$ and $P\cdot q=0$.

In the rest frame of the $Q\overline{Q}$ pair, which is our default choice,
the explicit forms of all the momenta are given by
\begin{subequations}
\bqa
P^\mu &=&(2E,0),\\
q^\mu &=&(0,{\bf q}),\\
p^\mu &=&(E,{\bf q}),\\
\bar{p}^\mu &=&(E,-{\bf q}).
\eqa
\label{Momenta:explicit:rest:frame}
\end{subequations}
Hence the total momentum $P$ becomes purely timelike, while the
relative momentum $q$ is purely spacelike, and one has $q^2=-{\bf
q}^2$ and $E=\sqrt{m^2+{\bf q}^2}$. In this frame, the momenta
carried by both photons have a magnitude of $E$.

For latter use, we define two velocity variables:
\begin{subequations}
\bqa
\beta &\equiv & {|{\bf q}|\over E},
\\
v & \equiv & {|{\bf q}|\over m}.
\eqa
\label{Momenta:explicit:rest:frame}
\end{subequations}
We will distinguish these two variables even in the nonrelativistic limit.

\subsection{Covariant projection approach}

\subsubsection{Projection of spin-singlet $Q\overline{Q}$ state}
\label{projection:spin-singlet}

We start with the quark amplitude $Q(p)\overline{Q}(\bar{p})\to
\gamma(k_1,\varepsilon_1) + \gamma(k_2,\varepsilon_2)$,
with the momenta of $Q$ and $\overline{Q}$ defined in (\ref{p-pbar-mom}):
\begin{equation}
\bar{u}(p) T v(\bar{p}) = \textrm{Tr} \big[ v(\bar{p})
\bar{u}(p) T \big].
\label{amplitud:QQbar}%
\end{equation}
Here $T$ denotes a matrix in Dirac-color space.

To proceed, we need first to project the amplitude
(\ref{amplitud:QQbar}) onto the spin-singlet color-singlet
$Q(p)\overline{Q}(\bar{p})$ state, by replacing the
$v(\bar{p})\bar{u}(p)$ with a suitable projection matrix. The
projector that is valid to all orders in $\mathbf{q}$ for the
spin-singlet color-singlet channel, denoted by
$\Pi_1^{(1)}(p,\bar{p})$, was first derived in \cite{Bodwin:2002hg}:
\bqa
\Pi_1^{(1)}(p,\bar{p})&=& \sum_{s_1,s_2} u(p,s_1) \bar{v}(\bar{p},s_2)
\langle {1\over 2},s_1;{1\over 2},s_2|00\rangle
\otimes  {\mathbf{1}_c\over \sqrt{N_c}}
\nn\\
&=& {1 \over 8\sqrt{2}E^2 (E + m)} (/\!\!\!{p}+m)(\,/\!\!\!P\!+\!2E )\,\gamma_5 (/\!\!\!\bar{p}-m)
\otimes  {\mathbf{1}_c\over \sqrt{N_c}},
\label{spin:singlet:projector}%
\eqa
where $\mathbf{1}_c$ is the unit matrix in the fundamental
representation of the color $SU(3)$ group.
The above spin-singlet projector is derived by
assuming the nonrelativistic normalization convention for Dirac spinor.
Applying this projector to
(\ref{amplitud:QQbar}), one obtains the amplitude for a
spin/color-singlet $Q\overline{Q}$ pair annihilation decay into two photons:
\beq
 {\mathcal A}^{\rm sing}[Q\overline{Q}\to \gamma\gamma] =
\textrm{Tr}\bigg\{ \Pi_1^{(1)}(p,\bar{p}) T \bigg\},
\label{spin:singlet:ampltude}%
\eeq
where the trace is understood to act on both Dirac and color spaces.

\subsubsection{Projection of $S$-wave amplitude}

In the amplitude in (\ref{spin:singlet:ampltude}), the $Q\overline{Q}$ pair is warranted to be in the
spin-singlet state, but not necessarily in the $S$-wave orbital-angular-momentum state. To
project out the $S$-wave amplitude, one needs average the amplitude
${\mathcal A}^{\rm sing}$ over all the directions of the relative
momentum $\mathbf{q}$ in the $Q\overline{Q}$ rest frame.
Afterwards one expands the resulting $S$-wave amplitude in powers of ${\bf q}^2$ and
truncate the series at the desired order.

A question may concern us immediately-- should the relative momentum $q$ be taken as
a 4-dimensional or a $D$-dimensional vector upon $S$-wave angular averaging?
Since for our process the matching can be conducted solely at amplitude level,
there is no {\it a priori} criterion to tell which treatment is superior.
At first glancing, our incapability to pick up a ``unique" and ``correct" scheme
may look troublesome, since different treatments may, conceivably, lead to
different answers for the full QCD amplitude once beyond LO in ${\bf q}^2$ expansion.
Nevertheless, it is possible that both schemes are equally acceptable,
provided that ultimately they yield {\it identical} short-distance coefficients.
In this respect, the ambiguity about the dimensionality of $q$
may turn into a virtue, in that it can serve as a useful consistency check
against our calculation.
In Section~\ref{sec:full:QCD:NLO}, we will separately treat $q^\mu$ to be 4-dimensional (dubbed $q_4$ scheme),
and $D$-dimensional (labeled $q_D$ scheme)~\footnote{Since $q^0=0$ in the rest frame of the $Q\overline{Q}$ pair,
it seems more natural to view $q^\mu$ as either a $3$-dimensional or a $D-1$-dimensional vector.
Nevertheless, in the lack of confusion, we will stick to the terms $q_4$ and $q_D$ scheme.}.

Since we are only concerned with the first-order relativistic correction,
we will follow a standard shortcut to extract the $S$-wave amplitude
rather than literally perform the angular integration.
In $q_4$ scheme, we first expand the spin-singlet amplitude ${\mathcal A}^{\rm sing}$ in $q^\mu$ through the
quadratic order, then make the following replacement:
\bqa
q^\mu q^\nu  &\to&  {\mathbf{q}^2 \over 3}\: \Pi^{\mu\nu}(P)\qquad\qquad(q_4\;{\rm scheme}),
\label{S-wave:projection:q4:scheme}%
\eqa
where
\bqa
\Pi^{\mu\nu}(P) &\equiv& -g^{\mu\nu}+{P^\mu P^\nu\over P^2}.
\nn %
\eqa
Subsequently we would be able to identify ${\mathscr A}_i$ ($i=0,2$)
as indicated in (\ref{full:amplitude:S:wave:series}).

If $q^\mu$ is assumed to be a $D$-dimensional Lorentz vector, one can follow the same step as described above,
except one make the following substitution:
\bqa
q^\mu q^\nu  &\to&  {\mathbf{q}^2 \over D-1}\: \Pi^{\mu\nu}(P)\qquad\qquad(q_D\;{\rm scheme}).
\label{S-wave:projection:qD:scheme}%
\eqa

We stress that in both schemes, the momenta $P$, $k_1$, $k_2$ and
polarization vectors $\varepsilon_1$, $\varepsilon_2$ are always
assumed to reside only in physical spacetime dimensions.

\subsection{$\bm{\gamma}_{\bm 5}$-prescription in Dimensional Regularization}

When applying the covariant spin-singlet projector as given in Sec.~\ref{projection:spin-singlet},
one needs to deal with the trace of $\gamma_5$ with a string of Dirac $\gamma$-matrices.
This will pose one notorious problem,
that a definite prescription of $\gamma_5$ must be specified if the spacetime dimension $D$ is
deformed from four.

We first point out that, for our process, in principle there exists no any technical subtlety
about $\gamma_5$ in the $q_4$ scheme. One can always choose to first calculate the quark amplitude
$Q\overline{Q}\to \gamma\gamma$ through NLO in $\alpha_s$.
As usual, dimensional regularization (DR) can be chosen to regularize both
ultraviolet (UV) and infrared (IR) divergences. After the loop integration is done,
one will end up with the $T$-matrix in (\ref{amplitud:QQbar})
that only depends on the external kinematic variables,
{\it e.g.} the momenta of quarks and photons, as well as the polarization vectors of photons,
which are all 4-dimensional objects.
Upon projecting out the spin-singlet amplitude, it is obviously legitimate to
use the standard 4-dimensional trace formula involving $\gamma_5$ in (\ref{spin:singlet:ampltude}).

In the $q_D$ scheme, the $\gamma_5$ problem cannot be circumvented even if one first carries out
the loop integration for the quark amplitude, because
the quark momenta $p$ and $\bar{p}$ appearing in the spin-singlet projector
(\ref{spin:singlet:projector}) can now penetrate into unphysical dimensions.
In this case, the rule about $D$-dimensional trace operation involving $\gamma_5$
in (\ref{spin:singlet:ampltude}) must be specified.

In practical computation, it is simpler to apply (\ref{spin:singlet:ampltude})
prior to carrying out the loop integration. Since the internal fermion propagators and vertices
are all $D$-dimensional objects, and all of them will enter into the trace,
it becomes compulsory to specify the prescription of $\gamma_5$ in DR,
for both $q_D$ and $q_4$ schemes.

In literature, there are two popular prescriptions about $\gamma_5$ in DR, the naive dimensional
regularization (NDR)~\cite{Chanowitz:1979zu,Korner:1991sx} and 't~Hooft-Veltman dimensional
regularization (HVDR)~\cite{'tHooft:1972fi,Breitenlohner:1977hr}.
In the former prescription, one assumes $\{\gamma_5,\gamma^\mu\}=0$ for all
$\mu=0,1,\cdots,D-1$. In the latter, one explicitly constructs $\gamma_5 \equiv i\gamma^0\gamma^1\gamma^2\gamma^3$,
which anticommutes with $\gamma^\mu$ for $\mu=0,1,2,3$
but commutes with $\gamma^\mu$ for $\mu=4,\cdots,D-1$.

In an arbitrary dimension, the definition ${\rm Tr}[\gamma_5\gamma^\mu\gamma^\nu\gamma^\alpha\gamma^\beta]=
-4i\epsilon^{\mu\nu\alpha\beta}$ and $\{\gamma_5,\gamma^\mu\}=0$ are incompatible~\cite{Breitenlohner:1977hr},
therefore $\gamma_5$ in NDR is an ambiguous object.
In order to obtain consistent predictions in this scheme, one must impose some additional rules,
{\it e.g.}, to give up the cyclicity property of trace and to place $\gamma_5$ in a fixed position
called ``reading point"~\cite{Korner:1991sx}.
Nevertheless, for its technical simplicity, the NDR scheme has been widely utilized
in computing the NLO QCD corrections to quarkonium decay and production processes~\cite{Petrelli:1997ge},
though most of which are at the LO accuracy in $v$ only~\footnote{
It has also been pointed out in \cite{Petrelli:1997ge}, that the $\gamma_5$-prescription implicit in
the threshold expansion technique developed
by Braaten and Chen~\cite{Braaten:1996jt,Braaten:1996rp} is essentially equivalent to NDR.}.

In contrast to NDR, the HVDR scheme turns to be a mathematically consistent scheme,
in which $\gamma_5$ is a well-defined and unique object. For example, the HVDR scheme
can automatically guarantee to recover the celebrated Adler-Bell-Jackiw chiral anomaly,
while in the NDR scheme, some {\it ad hoc} prescription has to be imposed to achieve this.

In HVDR, the $\gamma$-matrices in $D$ dimension obey the following anticommutation algebra:
\bqa
\{\gamma^\mu,\gamma^\nu\}&=& 2 g^{\mu\nu}, \quad\quad \{\gamma_5,\gamma^\mu\}=
2 \hat{g}^\mu_\nu \gamma_5 \gamma^\nu.
\label{gamma5:HVDR}
\eqa
$\hat{g}_{\mu\nu}$ denotes the projection of the metric tensor onto the unphysical dimensions,
which equals $g_{\mu\nu}$ for $\mu, \nu=4, \cdots, D-1$, and equals $0$ otherwise.
Some useful relations about this tensor are
$\hat{g}^\mu_\mu=D-4$, $\hat{g}_{\mu\alpha}\hat{g}^\alpha_\nu=\hat{g}_{\mu\alpha} g^\alpha_\nu= \hat{g}_{\mu\nu}$.
A nuisance of the HVDR scheme is that, since the first four dimensions are singled out
as special, Lorentz covariance has been sacrificed in the intermediate stage.
Moreover, one may get the impression that the messy anticommutation rule for $\gamma_5$ in $D$ dimension
would render the practical calculation a formidable task.

In this work, in favor of its internal consistency, we choose to work with the HVDR scheme.
It is necessary to spell out the recipe of $D$-dimensional trace operation involving $\gamma_5$ in this scheme.
In arbitrary spacetime dimension,
the trace of a $\gamma_5$ with odd number of $\gamma$-matrices
always vanishes. Starting from (\ref{gamma5:HVDR}), West has derived
a recursive formula for the trace of $\gamma_5$ with an even number of
$\gamma$-matrices~\cite{West:1991xv}:
\begin{subequations}
\bqa
& & {\rm Tr}[\gamma_5\gamma^{\mu_1}\gamma^{\mu_2}\gamma^{\mu_3}\gamma^{\mu_4}]=
-4 i\,\epsilon^{\mu_1\mu_2\mu_3\mu_4},
\\
&&{\rm Tr}[\gamma_5\gamma^{\mu_1}\gamma^{\mu_2}\cdots\gamma^{\mu_n}]=
\frac{2}{n-4}\sum_{i=2}^n\sum_{j=1}^{i-1}
(-1)^{i+j+1}g^{{\mu_i}{\mu_j}}{\rm Tr}\left[\gamma_5 \prod_{
k=1 (\neq i, j)}^n\gamma^{\mu_k}\right]\quad ({\rm for}
\;n\ge 6),
\nn\\
\eqa
\label{West:trace:gamma5:formula}
\end{subequations}
where the Levi-Civita tensor is a 4-dimensional object in HVDR, {\it i.e.},
$\hat{g}^\mu_\alpha \epsilon_{\mu\nu\beta\gamma}=0$.

One attractive point of West's trace formula is that, it involves only the 4-dimensional
antisymmetric tensor together with the $D$-dimensional metric tensor
(not the evanescent metric tensor $\hat{g}_{\mu\nu}$!). As a consequence,
this recursive algorithm can be readily implemented in the
\textsc{Mathematica} packages specialized to high energy physics, such as \textsc{FeynCalc}~\cite{Mertig:1990an}.

Equation~(\ref{West:trace:gamma5:formula}) is valid in any dimension, of course also in $D=4$, however it is
superficially much more involved than the familiar 4-dimensional trace formula~\footnote{For example, in the HVDR scheme,
the trace of $\gamma_5$ with six $\gamma$-matrices will result in 15 terms,
in contrast to the 6 terms that one would directly obtain in 4 dimension.}.

To make use of West's formula
when projecting out the spin-singlet amplitude, we can employ the cyclicity of trace
to move $\gamma_5$ in (\ref{spin:singlet:ampltude}) to the leftmost,
since this property persists to be a valid operation in the HVDR scheme:
\beq
{\mathcal A}^{\rm sing}(Q\overline{Q}\to \gamma\gamma) =
{1 \over 8\sqrt{2 N_c}E^2 (E + m)}
\textrm{Tr}\bigg\{\gamma_5 (/\!\!\!\bar{p}-m) T\,(/\!\!\!{p}+m)(\,/\!\!\!P\!+\!2E )
\bigg\},
\label{spin:singlet:ampltude:cyclicity}
\eeq
and the color trace has been implicit.

For obvious reason, we would not bother to use the anti-commutation relation
of $\gamma_5$ to realize this goal.

\section{QCD amplitude of $\bm{Q}\overline{\bm Q}\bm{(}{}^{\bm 1}\bm{S}_{\bm 0}^{\bm{[}\bm{1}\bm{]}}
\bm{)} \bm{\to} \bm{\gamma}\bm{\gamma}$ through NLO in $\bfalpha_{\bm s}$}
\label{sec:full:QCD:NLO}

In this section we employ the covariant projection technique
described in the preceding section to compute
$Q\overline{Q}({}^1S_0^{[1]})\to \gamma\gamma$ through NLO in
$\alpha_s$. The HVDR scheme will be used throughout.

\subsection{Tree-level amplitude and matching coefficients}
\label{tree:amplitude:matching}

\begin{figure}[tb]
\begin{center}
\includegraphics[height=6.0cm]{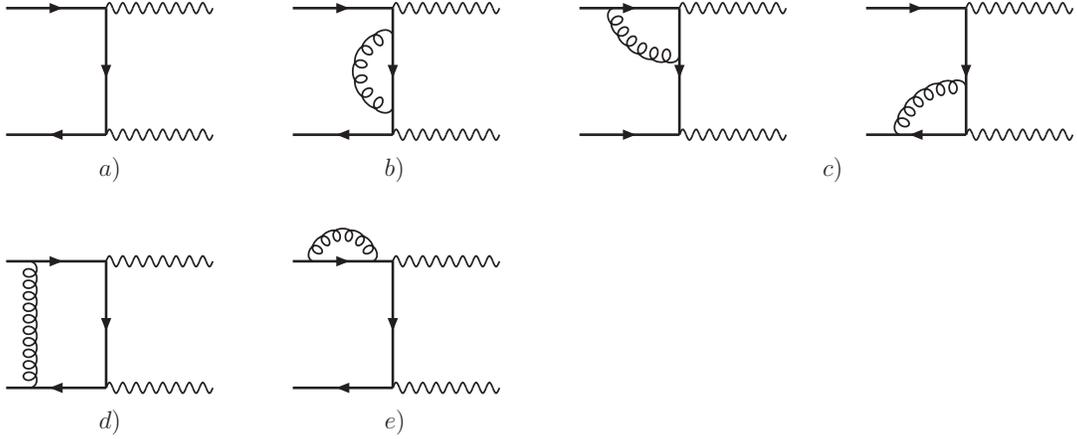}
\caption{Feynman diagrams for $Q\overline{Q}({}^1S_0^{[1]})
\to\gamma\gamma$ through ${\cal O}(\alpha_s)$. For simplicity, the
crossed diagrams have been suppressed.
\label{QCD:diag:etaQ:two:photon}}
\end{center}
\end{figure}

There are two ${\cal O}(\alpha_s^0)$ diagrams for $Q\overline{Q}\to
\gamma\gamma$, one of which is illustrated in
Fig.~\ref{QCD:diag:etaQ:two:photon}$a$). The corresponding LO
$T$-matrix reads:
\bqa
T^{(0)}&=& -i e^2 e_Q^2 \left[\not\!\varepsilon_2^* {
\not\!p-\not\!k_1+m\over -2 p\cdot k_1}\not\!\varepsilon_1^*+
\not\!\varepsilon_1^* {-\!\not\!\bar{p}\,+\not\!k_1+m\over -2
\bar{p}\cdot k_1 }\not\!\varepsilon_2^* \right] \otimes
\mathbf{1}_c. \label{tree:amplitude}
\eqa

If $q$ lives only in physical dimensions, one can directly substitute (\ref{tree:amplitude})
into (\ref{spin:singlet:ampltude}) to project out the spin-singlet amplitude,
and use the 4-dimensional trace formula to obtain~\cite{Bodwin:2002hg}:
\bqa
\mathcal{A}^{\rm sing(0)}(Q\overline{Q}\to \gamma\gamma)
&=&  e^2 e_Q^2 \sqrt{2N_c} \,\hat{\mathbf k}_1 \cdot \bfvarepsilon_1^* \times \bfvarepsilon_2^*
{m E^2\over E^4-(k_1\cdot q)^2}.
\label{spin-singlet:ampltude:tree:q4}%
\eqa

If $q$ is instead allowed to leak into the unphysical dimensions,
one needs substitute (\ref{tree:amplitude}) into
(\ref{spin:singlet:ampltude:cyclicity}), and utilize
(\ref{West:trace:gamma5:formula}) to carry out the $D$-dimensional
trace:
\bqa
& & {\mathcal A}^{\rm sing(0)}(Q\overline{Q}\to \gamma\gamma) = e^2 e_Q^2 \sqrt{N_c\over 2} {1 \over (E+m)(E^4-(k_1\cdot q)^2)}
\nn\\
&\times & \bigg[E(E+m) \epsilon^{\mu\nu\alpha\beta} k_{1\alpha}
+ \epsilon^{\mu \alpha \beta \gamma} k_{1\alpha} q_\gamma q^\nu - \epsilon^{\nu \alpha \beta \gamma} k_{1\alpha} q_\gamma q^\mu
 + \epsilon^{\mu \nu \alpha \beta} q_\alpha k_1\cdot q
 \bigg]P_\beta \varepsilon_{1\mu}^* \varepsilon_{2\nu}^*.
\label{spin-singlet:ampltude:tree:qD}%
\eqa
We have dropped those terms linear in $q$, which trivially vanish
due to the constraints $P\cdot \varepsilon_i=0$ and $P^0=2 k_1^0$.
Note this expression is much more complicated than
(\ref{spin-singlet:ampltude:tree:q4}), though they should be exactly
identical when $q$ is a 4-dimensional vector.

The Levi-Civita tensor, $k_1$, $\varepsilon_1$ and $\varepsilon_2$
are all 4-dimensional quantities. For any $q$ vector in
(\ref{spin-singlet:ampltude:tree:qD}) that contracts with them, only
its first four components can contribute, that is, one can make the
replacement $q^\alpha\to \bar{q}^\alpha \equiv
(g^\alpha_\beta-\hat{g}^\alpha_\beta) q_\beta$, without affecting
the answer. The unphysical components of $q$, $\hat{q}^\alpha \equiv
\hat{g}^\alpha_\beta q_\beta$, contribute to
(\ref{spin-singlet:ampltude:tree:qD}) only implicitly through the
factor $E$.

Expand (\ref{spin-singlet:ampltude:tree:q4}) and (\ref{spin-singlet:ampltude:tree:qD})
in $q$, then apply (\ref{S-wave:projection:q4:scheme}) and (\ref{S-wave:projection:qD:scheme}),
one can pick up the corresponding tree-level $S$-wave amplitudes in the $q_4$ and $q_D$ schemes.
It is then straightforward to identify ${\mathscr A}^{(0)}_i$ as introduced
in (\ref{full:amplitude:S:wave:series}):
\begin{subequations}
\bqa
& & {\mathscr A}^{(0)}_{0}=\sqrt{2N_c}\,{4\pi
e_Q^2\alpha  \over m},
\\
& & {\mathscr A}^{{(0)}}_2\bigg|_{q_4\;{\rm scheme}}= -\sqrt{2N_c}\,{8 \pi
e_Q^2\alpha  \over 3 m},
\\
& & {\mathscr A}^{{(0)}}_2\bigg|_{q_D\;{\rm scheme}}= -\sqrt{2N_c}\,{2 D \pi
e_Q^2\alpha  \over (D-1) m}.
\eqa
\label{mathscr:A:LO:alphas}
\end{subequations}
Not surprisingly, the ${\cal O}(v^2)$ $S$-wave amplitudes differ in
$q_4$ and $q_D$ schemes.

The NRQCD matarix elements at LO in $\alpha_s$ have been given in
(\ref{NRQCD:matrix:elements:LO}). According to (\ref{QQbar:2gamma:ampl:level:NRQCD:ampl}),
one can write down the Born-order
perturbative NRQCD amplitude:
\beq
{\mathbb A}_{\rm NRQCD}^{(0)}=\sqrt{2N_c} \left[ c_0^{(0)}+c_2^{(0)} v^2 +\cdots \right].
\label{Ampl:NRQCD:tree}
\eeq

Combining (\ref{Matching:formula:ampl:QQbar:2gamma}), (\ref{full:amplitude:S:wave:series}),
(\ref{mathscr:A:LO:alphas}) and (\ref{Ampl:NRQCD:tree}),
one easily recognizes the  ${\cal O}(\alpha_s^0)$ short-distance coefficients
$c_i^{(0)}= {\mathscr A}_i^{{(0)}}/\sqrt{2N_c}$ ($i=0,2$):
\begin{subequations}
\bqa
& & c_{0}^{(0)} = {4\pi e_Q^2 \alpha \over m},
\label{c0:tree}
\\
& & c_2^{(0)}\bigg|_{q_4\;{\rm scheme}} = -{8\pi e_Q^2\alpha \over 3 m},
\label{c2:tree:q4}
\\
& & c_2^{(0)}\bigg|_{q_D\;{\rm scheme}} = -{2-\epsilon\over 3-2\epsilon}{4\pi e_Q^2\alpha\over m},
\label{c2:tree:qD}
\eqa
\label{c0:and:c2:tree}
\end{subequations}
where as usual, $D\equiv 4-2\epsilon$.
The coefficient $c_2^{(0)}$ in $q_D$ scheme differs from that in $q_4$ scheme
by an ${\cal O}(\epsilon)$ constant.
At first sight, it seems no need to retain this extra piece
since no any divergence emerges at this order.
However, as will become clear later,
this ${\cal O}(\epsilon)$ piece plays a key role for ultimately obtaining
the scheme-independent ${\cal O}(\alpha_s)$ short-distance coefficients.

\subsection{QCD amplitude at NLO in $\bfalpha_{\bm s}$}
\label{QCD:ampl:NLO:alphas}

We proceed to compute NLO QCD correction to the
$Q\overline{Q}({}^1S_0^{[1]})\to \gamma\gamma$ process. At ${\cal
O}(\alpha_s)$, there are eight one-loop diagrams, including two
self-energy diagrams, four triangle diagrams and two box diagrams.
Half of these diagrams have been shown in
Fig.~\ref{QCD:diag:etaQ:two:photon}$b$) through
\ref{QCD:diag:etaQ:two:photon}$d$).

Each individual NLO diagram may contain UV or IR divergences. For example, the self-energy
and triangle diagrams contain UV divergences, and the box diagrams possess IR divergence.
We will choose DR as a convenient regulator to regularize both types of divergences.
Since our concern is to calculate the gauge-invariant on-shell amplitude, for simplicity
we will work with Feynman gauge in this Section.

In accordance with the LSZ reduction formula, we need multiply the tree-level amplitude in
Fig.~\ref{QCD:diag:etaQ:two:photon}$a$)
by the residue of the heavy quark propagator at its pole, $Z_Q$.
This contribution is represented by Fig.~\ref{QCD:diag:etaQ:two:photon}$e$).
In Feynman gauge, the residue is given by
\bqa
Z_Q &=& 1 - {C_F \alpha_s \over 4\pi}
\left(\frac{1}{\epsilon_{\rm UV}}+\frac{2}{\epsilon_{\rm IR}}
-3\gamma_E+3\ln\frac{4\pi\mu^2}{m^2}+4\right)+{\cal
O}(\alpha_s^2),
\label{Z-Q}
\eqa
where $\gamma_E$ is the Euler constant, and $C_F= {N_c^2-1\over 2 N_C}$ is the Casmir for the
fundamental representation of the $SU(N_c)$ group.

In addition, we also need replace the bare quark mass in the quark propagator in
Fig.~\ref{QCD:diag:etaQ:two:photon}$a$) by
\bqa
m^{\rm bare} &=& m\bigg[1 - {C_F \alpha_s \over 4\pi} \left(\frac{3}{\epsilon_{\rm
UV}}-3\gamma_E+3\ln\frac{4\pi\mu^2}{m^2}+4\right)+{\cal
O}(\alpha_s^2)\bigg]\,.
\label{mass:renormalization}
\eqa

It is straightforward to write down the respective $T^{(1)}$-matrix for each NLO
diagram in Fig.~\ref{QCD:diag:etaQ:two:photon}.
We then substitute them into (\ref{spin:singlet:ampltude:cyclicity}), and for simplicity,
use (\ref{West:trace:gamma5:formula}) to carry out the
$D$-dimensional trace prior to performing loop integration~\footnote{Since
$\gamma_5$ in the HVDR scheme is mathematically unambiguous,
reversing the order of trace and loop integration would not affect
the final result.}.

Practically, we resort to the \textsc{Mathematica} package \textsc{FeynCalc}~\cite{Mertig:1990an}
to accomplish the abovementioned trace operation,
because West's formula (\ref{West:trace:gamma5:formula}) is its built-in algorithm for
calculating the trace involving $\gamma_5$.
We continue to use \textsc{FeynCalc} to reduce all the encountered one-loop tensor integrals to
the one-loop scalar integrals. It turns out that only a couple of
two-point, three-point scalar integrals and one four-point scalar integral are required.
For reader's convenience,
the closed-form expressions for these scalar integrals
have been collected in Appendix~\ref{appendix}.

By far we have obtained the analytic expression for
$\mathcal{A}^{\rm sing(1)}$, the ${\cal O}(\alpha_s)$ spin-singlet
amplitude for $Q\overline{Q}\to \gamma\gamma$. We proceed to expand
it to second order in $q$, then apply
(\ref{S-wave:projection:q4:scheme}) and
(\ref{S-wave:projection:qD:scheme}) to extract the corresponding
$S$-wave amplitudes defined in (\ref{full:amplitude:S:wave:series}),
in both $q_4$ and $q_D$ schemes, respectively. The intermediate
steps are straightforward but cumbersome, and some special care has
to be paid when dealing with the box diagrams. Fortunately, almost
all these manipulations can be handled by computer.

\begin{table}
\caption{The individual contributions to the
$Q\overline{Q}(^1S_0^{[1]})\to 2 \gamma$ amplitude, ${\mathscr A}_0$
and ${\mathscr A}_2$, from different classes of diagrams in
Fig.~\ref{QCD:diag:etaQ:two:photon}. A common factor
$\sqrt{2N_c}\left({4\pi e_Q^2\alpha\over m}\right)$ has been
suppressed for the Born-order results, while a common factor
$\sqrt{2N_c}\left({4 C_F e_Q^2 \alpha\alpha_s\over m}\right)$ has
also been dropped for the ${\cal O}(\alpha_s)$ results. For brevity,
we have used the shorthand
${1\over\hat{\epsilon}}\equiv{1\over\epsilon}-\gamma_E+\ln{4\pi\mu^2
\over m^2}$. \label{Table:QCD:Amplitude}}
\begin{ruledtabular}
\begin{tabular}{ccc}
QCD diagrams & $q_4$ scheme & $q_D$ scheme
\\
\hline
 & ${\mathscr A}_0$ &
\\
\hline
\ref{QCD:diag:etaQ:two:photon}$a$) & \multicolumn{2}{c}{1}
\\
\ref{QCD:diag:etaQ:two:photon}$b$)  &\multicolumn{2}{c}{$-{1\over
\hat{\epsilon}_{\rm UV}}+2\ln2-\frac{3}{2}$}
\\
\ref{QCD:diag:etaQ:two:photon}$c$)
&\multicolumn{2}{c}{$\frac{1}{2}{1\over\hat {\epsilon}_{\rm
UV}}-2\ln2+\frac{\pi^2}{8}$}
\\
\ref{QCD:diag:etaQ:two:photon}$d$) &\multicolumn{2}{c}{${\pi^2\over
4v}+ {i\,\pi\over 4 v} \left( -{1\over \hat{\epsilon}_{\rm IR}}+2
\ln(2v)\right) + {1\over 2}{1\over\hat {\epsilon}_{\rm IR}}-1$}
\\
\ref{QCD:diag:etaQ:two:photon}$e$) &\multicolumn{2}{c}{${1\over 2}
\left({1\over\hat {\epsilon}_{\rm UV}}-{1\over\hat {\epsilon}_{\rm
IR}}\right)$}
\\
\hline
&  ${\mathscr A}_2$ &
\\
\hline
\ref{QCD:diag:etaQ:two:photon}$a$) &  $-{2\over 3}$ &   $ -{2 \over
3} - {\epsilon \over 9}$
\\
\ref{QCD:diag:etaQ:two:photon}$b$) &{$\frac{7}{6}{1\over
\hat{\epsilon}_{\rm
UV}}-\frac{8}{3}\ln2+\frac{7}{3}$}&{$\frac{7}{6}{1\over
\hat{\epsilon}_{\rm UV}}-\frac{8}{3}\ln2+\frac{41}{18}$}
\\
\ref{QCD:diag:etaQ:two:photon}$c$) &$-{1 \over 3}{1\over\hat
{\epsilon}_{\rm UV}}+ {10\over 3}\ln2- {\pi^2 \over 8}- {1\over 6}$&
$-{1 \over 3}{1\over\hat {\epsilon}_{\rm UV}}+ {10\over 3}\ln2-
{\pi^2 \over 8}- {2\over 9}$
\\
\ref{QCD:diag:etaQ:two:photon}$d$) & $ {5\pi^2 \over 24 v}+
{5i\,\pi\over 24v}\left( -{1\over\hat{\epsilon}_{\rm IR}}+
2\ln(2v)\right)$ &$ {5\pi^2 \over 24v} + {5i\,\pi\over 24v}\left(
-{1\over\hat{\epsilon}_{\rm IR}}+ 2\ln(2v) +{2\over15} \right)$
\\
&  $+ {1 \over 3}{1\over
\hat{\epsilon}_{\rm IR}}-2\ln{2}-{4 \over 9}$
&  $+ {1\over 3}{1\over
\hat{\epsilon}_{\rm IR}}-2\ln{2}- {1\over 2}$
\\
\ref{QCD:diag:etaQ:two:photon}$e$) & $-\frac{5}{6}{1\over
\hat{\epsilon}_{\rm UV}}+ {1 \over 3}{1\over \hat{\epsilon}_{\rm
IR}}- {2 \over 3}$ & $-\frac{5}{6}{1\over \hat{\epsilon}_{\rm UV}}+
{1\over 3}{1\over \hat{\epsilon}_{\rm IR}}-{1\over 2}$
\\
\end{tabular}
\end{ruledtabular}
\end{table}

For completeness and for clarity, we tabulate in
Table.~\ref{Table:QCD:Amplitude} the individual contributions to the
$S$-wave amplitudes ${\mathscr A}^{(1)}_{0}$ and ${\mathscr
A}^{(1)}_{2}$ from each class of diagrams, in both $q_4$ and $q_D$
schemes. For each entry, we sum all the diagrams that belong to the
same topology class, including the crossed ones. The entry
``\ref{QCD:diag:etaQ:two:photon}$e$)" in
Table~\ref{Table:QCD:Amplitude} refers to the ${\cal O}(\alpha_s)$
contributions from both wave function and mass renormalization, as
specified in (\ref{Z-Q}) and (\ref{mass:renormalization}).

As can be seen from Table~\ref{Table:QCD:Amplitude}, the Coulomb
singularity arises in the box diagrams, present in both ${\mathscr
A}^{(1)}_{0}$ and ${\mathscr A}^{(1)}_{2}$. The emergence of this
$\pi^2/v$ singularity reflects that the real part can receive the
contribution from the {\it potential} region. For completeness, we
also explicitly include the contributions of the imaginary part from
the box diagrams. This corresponds to a configuration that the two
non-adjacent $Q$ and $\overline{Q}$ quarks in the box can become
simultaneously on their mass-shells, so that the exchanged gluon can
only be {\it potential} mode. Consequently, only the potential
region can contribute to the imaginary part, which is plagued with
the joint $i\pi/v$ and IR singularities as well as the term
nonanalytic in $\bf q$. We expect that the corresponding ${\cal
O}(\alpha_s)$ NRQCD calculation in next section will exactly
reproduce such potential-region contributions.

Another observation from Table.~\ref{Table:QCD:Amplitude} is that, both the $q_4$ and $q_D$ schemes
differ on the ${\cal O}(v^2)$ contributions in each individual diagram. This pattern is in sharp
contrast to the ${\cal O}(v^0)$ case.

Summing up all the individual contributions listed in
Table.~\ref{Table:QCD:Amplitude}, one finally obtains the complete
NLO QCD amplitude for $Q\overline{Q}({}^1S_0^{[1]})\to
\gamma\gamma$:
\begin{subequations}
\bqa
{\mathscr A}^{(1)}_{0}&=& \sqrt{2N_c}\,{4C_F e_Q^2\alpha
\alpha_s \over m} \left[{\pi^2-20\over 8}+ \frac{\pi^2}{4v}
+ {i\,\pi\over
4v}\left(-{1 \over \epsilon_{\rm IR}}+\gamma_E +\ln{{\bf q}^2\over \pi\mu^2}\right)\right],
\label{QCD:ampl:LO:v:NLO:alphas}
\\
{\mathscr A}^{(1)}_2 &=& \sqrt{2N_c}\, {4 C_F e_Q^2\alpha \alpha_s
\over m} \bigg[ \frac{2}{3}\left({1\over \epsilon_{\rm IR}}
-\gamma_E+ \ln{4\pi\mu^2
\over m^2}\right)+\frac{19}{18}-\frac{\pi^2}{8}-\frac{4}{3}\ln2\nn\\
&& + \,{5\pi^2 \over 24v}+ {5i\,\pi\over 24 v}
\left(-{1 \over \epsilon_{\rm IR}}+\gamma_E +\ln{{\bf q}^2\over \pi\mu^2} \right) + C_q\bigg].
\label{QCD:ampl:NLO:v:NLO:alphas}
\eqa
\label{QCD:ampls:NLO:alphas}
\end{subequations}
Interestingly, both the $q_4$ and $q_D$ schemes now fully agree on
the real part of the complete ${\cal O}(v^2)$ amplitude, and only
differ slightly on the imaginary part. The scheme dependence is
encoded in $C_q$, an imaginary constant, which equals $0$ in $q_4$
scheme, and equals ${i\pi \over 36v}$ in $q_D$ scheme. As will be
seen in Sec.~\ref{Matching:short-distance:NLO}, the
scheme-dependence of $C_q$ is intimately correlated with that of
$c_2^{(0)}$ in (\ref{c0:and:c2:tree}).

We make some further comments on the complete NLO QCD amplitude in (\ref{QCD:ampls:NLO:alphas}).
The UV divergences have been swept through the wave function and mass renormalization.
At LO in $v^2$, the IR divergences in the real part completely cancel when summing up all diagrams, which is
warranted by the neutral color charge of the $S$-wave $Q\overline{Q}$ pair~\cite{Bodwin:1994jh}.
Nevertheless, this cancelation fails to hold at NLO in $v^2$, and the presence of a net infrared divergence
in the real part of (\ref{QCD:ampl:NLO:v:NLO:alphas})
signals that the simple ``color-transparency" picture no longer applies,
and the soft gluons can resolve the geometric details of
the $Q\overline{Q}$ pair and may even strongly interact with its color dipole.
It is easy to find that this IR divergence arises from the {\it soft} region of the loop diagrams.
We anticipate that the corresponding ${\cal O}(\alpha_s)$ NRQCD calculation in next Section
will exactly reproduce this infrared divergence at relative order $v^2$.

\section{NRQCD amplitude at NLO in $\bfalpha_{\bm s}$}
\label{sec:NRQCD:NLO}

To match the accuracy of the QCD calculation for
$Q\overline{Q}({}^1S_0^{[1]})\to \gamma\gamma$ in
Section~\ref{QCD:ampl:NLO:alphas}, it is necessary to compute the
perturbative NRQCD amplitude ${\mathbb A}_{\rm NRQCD}$ to NLO in
$\alpha_s$.

In (\ref{NRQCD:matrix:elements:LO}), the two involved NRQCD matrix
elements have been given at LO in $\alpha_s$. In the following we
proceed to compute their ${\cal O}(\alpha_s)$ corrections. For this
purpose, the knowledge of the NRQCD lagrangian in the heavy quark
bilinear sector is required~\cite{Bodwin:1994jh}:
\bqa
{\mathcal L}_{\rm NRQCD} &=&
\psi^\dagger \left( i D_0 + {{\bf D}^2 \over 2m} \right) \psi+ \psi^\dagger {{\bf D}^4 \over 8m^3} \psi
+ {c_F \over 2 m} \psi^\dagger \bfsigma \cdot g_s {\bf B} \psi
\nn\\
&+& {c_D\over 8 m^2} \psi^\dagger ({\bf D}\cdot g_s {\bf E}- g_s {\bf E}\cdot {\bf D})\psi
+{i c_S\over 8 m^2} \psi^\dagger \bfsigma \cdot ({\bf D}\times g_s {\bf E}- g_s {\bf E}\times {\bf D})\psi
\nn\\
&+& \left(\psi \rightarrow i \sigma ^2 \chi^*, A_\mu \rightarrow - A_\mu^T\right) +
{\mathcal L}_{\rm light} \,.
\label{NRQCD:Lag}
\eqa
The replacement in the last line implies that the corresponding
heavy anti-quark bilinear sector can be obtained through the charge
conjugation transformation. ${\mathcal L}_{\rm light}$ represents
the lagrangian for the light quarks and gluons.

\begin{figure}[tb]
\begin{center}
\includegraphics[height=8.0 cm]{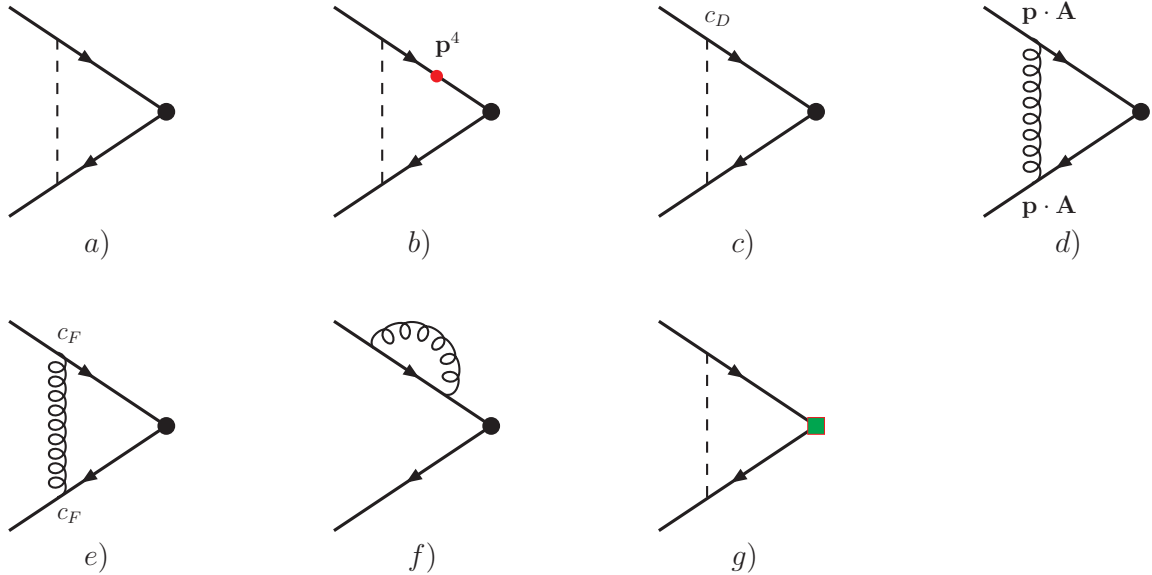} \caption{
The one-loop NRQCD diagrams that initiate the ${\cal O}(\alpha_s)$
corrections to the vacuum-to-$Q\overline{Q}({}^1S_0^{[1]})$ matrix
elements. The big solid circle represents the operator
$\chi^\dagger\psi$, while the solid square represents the operator
$\chi^\dagger(-\frac{i}{2}\tensor{\bold{D}})^2 \psi$. The dashed
line denotes the temporal gluon, and the curly line denotes the
transverse gluon. The unlabeled vertex attached to a temporal gluon
stands for the Coulomb interaction $\mp i g_s T^a$. For simplicity,
we have suppressed the conjugate diagrams of $b$) and $c$), in which
the same operators are inserted on the antiquark line.
\label{NRQCD-diagrams}}
\end{center}
\end{figure}

The short-distance coefficients $c_F$, $c_D$, $c_S$, in (\ref{NRQCD:Lag}) are associated
with the Fermi, Darwin, and spin-obit interactions, respectively.
To our purpose, suffices it to know their tree-level values
$c_F= c_D = c_S=1+{\cal O}(\alpha_s)$. Moreover, in (\ref{NRQCD:Lag}) we have neglected
all other higher-dimensional operators, since the relativistic corrections generated by
them are beyond the intended ${\cal O}(v^2)$ accuracy.

Through the relative order $v^2$, the ${\cal O}(\alpha_s)$ corrections to the
matrix elements stem from the one-loop NRQCD diagrams as illustrated in Fig.~\ref{NRQCD-diagrams}.
The first six diagrams~\footnote{We have not drawn the diagram in which a temporal gluon attaches to
 a spin-orbit interaction vertex and a Coulomb vertex, because its contribution simply
 vanishes due to spin conservation.} represent the one-loop correction to the matrix element
 $\langle0\vert\chi^\dagger\psi\vert Q\overline{Q}({}^1S_0^{[1]})\rangle$,
while the last one in Fig.~\ref{NRQCD-diagrams} constitutes the
one-loop correction to $\langle
0\vert\chi^\dagger(-\frac{i}{2}\tensor{\bold{D}})^2 \psi\vert
Q\overline{Q}({}^1S_0^{[1]})\rangle$.

In passing, we note that our one-loop correction calculation for the vacuum-to-$Q\overline{Q}({}^1S_0)$
NRQCD matrix elements is quite similar to the existing one for the vacuum-to-$Q\overline{Q}({}^3S_1)$
matrix elements~\cite{Luke:1997ys}.
Since heavy quark spin symmetry is violated at relative order $v^2$,
we cannot completely transplant their results here.
Ref.~\cite{Luke:1997ys} uses Feynman gauge in calculating NRQCD diagrams.
In contrast, in this section we will work in the Coulomb gauge,
since it has the virtue that the instantaneous (potential) nature of the temporal gluon becomes manifest.
Of course, insofar as the on-shell matrix element is concerned, the choice of gauge
is merely a matter of convenience.

When computing the loop diagrams in Fig.~\ref{NRQCD-diagrams}, we
first integrate over the temporal component $k^0$ using contour
integration, where $k$ signifies the loop momentum carried by the
gluon propagator. We then perform the remaining integration over
${\bf k}$ in $D-1$ spatial dimension. When the exchanged gluon is
transverse, as exemplified by Fig.~\ref{NRQCD-diagrams}$d$), the
situation is somewhat more complicated. Upon integrating over $k^0$,
one would end up with two terms, one with the residue taken at the
pole of the quark propagator, and the other with the residue taken
at the pole of the gluon propagator. The denominators of each
resulting term are in general inhomogeneous in powers of $v$. This
problem was initially overcome in a heuristic way by expanding the
integrand~\cite{Bodwin:1994jh}. This recipe was later clarified and
systemized in the {\it method of region}
formalism~\cite{Beneke:1997zp} (see also \cite{Luke:1997ys}). In
this context, the two terms after contour integration can be
identified with the potential and soft region, respectively. For
each region, loop momentum has a definite power in $v$ so that one
can readily homogenize the integrand by expanding the denominator
accordingly. After this is done, it then becomes a straightforward
exercise to carry out the spatial integrations in DR.

According to the LSZ reduction formula, we need multiply the LO
matrix element $\langle0\vert\chi^\dagger\psi\vert
Q\overline{Q}({}^1S_0)\rangle^{(0)}$ by the residue of the heavy
quark propagator in NRQCD near its pole, $Z_{\rm NRQCD}$. The ${\cal
O}(\alpha_s)$ piece of this contribution is represented by
Fig.~\ref{NRQCD-diagrams}$f$). In Coulomb gauge, the residue in DR
reads~\footnote{The Coulomb-gauge residue has been given in Appendix
B of \cite{Bodwin:1994jh}, which depends on a hard momentum cutoff
there. We have converted that cutoff-scheme expression to its
counterpart in DR.}
\bqa
{\label{Z:Coulomb:NRQCD}}
Z_{\rm NRQCD}(q)=1- {C_F \alpha_s \over 3\pi} v^2 \left( {1\over \epsilon_{\rm UV }}
- {1\over \epsilon_{\rm IR}}\right).
\eqa
It is worth noting that, the simultaneous occurrences of UV pole and IR pole at ${\cal O}(v^2)$ can
be traced to a logarithmically-divergent scaleless integral in the soft region.

\begin{table}
\caption{
The analytic expressions of each ${\cal O}(\alpha_s)$ NRQCD diagram in Fig.~\ref{NRQCD-diagrams} (in Coulomb gauge).
For simplicity, we have stripped
off an overall factor of $\sqrt{2N_c}{C_F\alpha_s\over \pi}$.
\label{Table:NRQCD:Amplitude}}
\begin{ruledtabular}
\begin{tabular}{clll}
NRQCD diagrams & & Expressions
\\
\hline
\ref{NRQCD-diagrams}$a$) && ${1\over 4 v} \left[\pi^2+ i\pi
\left(-{1\over \epsilon_{\rm IR}}+\gamma_E +\ln{{\bf q}^2\over
\pi\mu^2}-{v^2\over 4}\right)\right]$
\\
\ref{NRQCD-diagrams}$b$) &&  ${ v\over 8} \left[\pi^2+ i
\pi\left(-{1\over \epsilon_{\rm IR}}+\gamma_E+\ln{{\bf
q}^2\over\pi\mu^2}+{1\over 2}\right)\right]$
\\
\ref{NRQCD-diagrams}$c$)  &&  $-{v\over 4 } (i \pi)$
\\
\ref{NRQCD-diagrams}$d$) & &   ${v\over 4 }\left[\pi^2+ i \pi
 \left(-{1\over \epsilon_{\rm IR}}+\gamma_E + \ln{{\bf q}^2\over \pi\mu^2}-1\right)\right] -
 {v^2\over 3} \left({1\over \epsilon_{\rm UV}}-{1\over \epsilon_{\rm IR}}\right)$
\\
\ref{NRQCD-diagrams}$e$) & &  ${v\over 2 } (i \pi)$
\\
\ref{NRQCD-diagrams}$f$)  && $-{v^2\over 3} \left({1\over
\epsilon_{\rm UV}}-{1\over \epsilon_{\rm IR}}\right)$
\\
\hline
\ref{NRQCD-diagrams}$g$) && $m^2 {v\over 4 } \left[\pi^2 + i \pi
\left(-{1\over \epsilon_{\rm IR}}+\gamma_E + \ln{{\bf q}^2\over
\pi\mu^2}\right)\right]$
\\
\hline
\end{tabular}
\end{ruledtabular}
\end{table}

For clarity, we list in Table~\ref{Table:NRQCD:Amplitude} the
separate contribution from each diagram in
Fig.~\ref{NRQCD-diagrams}. In Fig.~\ref{NRQCD-diagrams}$a$), we
follow \cite{Luke:1997ys} to include an extra source of relativistic
correction that arises form the nonrelativistic expansion of the
kinematic energy of the external (anti)quark ${{\bf q}^2 \over
2m}-{{\bf q}^4\over 8m^3}$. For the entries corresponding to
Fig.~\ref{NRQCD-diagrams}$b$) and $c$), we have also included the
contributions from their conjugate diagrams. It is clear to see
that, all the diagrams receive contributions from the potential
region (odd powers of $v$). In addition,
Fig.~\ref{NRQCD-diagrams}$d$) and $f$) also receive the nonvanishing
contributions from the soft region (even powers of $v$), which
result in a logarithmically divergent scaleless integral. Only ${\bf
p}\cdot {\bf A}$ interaction can result in such a contribution.

Summing all the individual contributions in Table~\ref{Table:NRQCD:Amplitude}, we obtain the
${\cal O}(\alpha_s)$ corrections to the perturbative NRQCD matrix elements:
\begin{subequations}
{\label{NRQCD:ME:NLO}}
\bqa
& & \langle0\vert\chi^\dagger\psi\vert Q\overline{Q}({}^1S_0^{[1]})
\rangle^{(1)}
=\sqrt{2N_c} {C_F\alpha_s\over \pi}
\Bigg\{{1\over 4v}\left(1+{3 \over 2}v^2\right)
\bigg[\pi^2+i\pi \left(-{1 \over \epsilon_{\rm IR}}+\gamma_E +\ln{{\bf q}^2\over \pi\mu^2}
\right)\bigg]
\nn\\
& & \qquad\qquad \qquad -\frac{2v^2}{3}\bigg(\frac{1}{\epsilon_{\rm UV}}-
\frac{1}{\epsilon_{\rm IR}}\bigg)\Bigg\},
\label{NRQCD:ME:v0:oneloop}
\\
& & \langle 0\vert\chi^\dagger(-\frac{i}{2}\tensor{\bold{D}})^2
\psi\vert Q\overline{Q}({}^1S_0^{[1]})\rangle^{(1)}
= \sqrt{2N_c} { C_F\alpha_s\over \pi} m^2 \left({v \over 4}\right)
\bigg[\pi^2+i\pi \left(-{1 \over \epsilon_{\rm IR}}+\gamma_E +\ln{{\bf q}^2\over \pi\mu^2}
\right)\bigg].
\nn\\
\label{NRQCD:ME:v2:oneloop}
\eqa
\end{subequations}
As is evident in (\ref{NRQCD:ME:v0:oneloop}), the order-$\alpha_s$ correction to the
matrix element $\langle 0\vert\chi^\dagger\psi\vert Q\overline{Q}({}^1S_0)
\rangle$ is logarithmically UV divergent, which can be traced back to the
${\bf p}\cdot {\bf A}$ interactions in Fig.~\ref{NRQCD-diagrams}$d$) and $f$).
Employing the $\overline{\rm MS}$ scheme to subtract the UV divergence, we then obtain
the renormalized matrix element
\bqa
\langle 0\vert\chi^\dagger\psi\vert Q\overline{Q}({}^1S_0^{[1]})
\rangle^{(1)}_{\overline{\rm MS}} &=& \sqrt{2N_c} {C_F\alpha_s\over
\pi} \Bigg\{{1\over 4v}\left(1+{3 \over 2}v^2\right)
\bigg[\pi^2+i\pi \left(-{1 \over \epsilon_{\rm IR}}+\gamma_E
+\ln{{\bf q}^2\over \pi\mu^2} \right)\bigg]
\nn\\
&& + {2v^2 \over 3}\left({1 \over \epsilon_{\rm IR}} - \gamma_E + \ln 4\pi \right)\Bigg\}.
\label{MSbar:renormalized:v0:NRQCD:ME}
\eqa

According to (\ref{QQbar:2gamma:ampl:level:NRQCD:ampl}), accurate up to the order $\alpha_s v^2$,
we can express the perturbative NRQCD amplitude as
\bqa
& & {\mathbb A}_{\rm NRQCD} =
 (c_0^{(0)}+c_0^{(1)}) \langle 0| \chi^\dagger \psi |Q\overline{Q}({}^1S_0^{[1]})\rangle^{(0)}
+ c_0^{(0)} \langle 0\vert\chi^\dagger\psi\vert
Q\overline{Q}({}^1S_0^{[1]}) \rangle^{(1)}_{\overline{\rm MS}}
\label{NRQCD:amp:everything}
\\
 && + {c_2^{(0)}+c_2^{(1)} \over m^2}
  \langle 0| \chi^\dagger
(-\tfrac{i}{2}\tensor{\mathbf{D}})^{2}\psi|
Q\overline{Q}({}^1S_0^{[1]})\rangle^{(0)} + {c_2^{(0)} \over m^2}
\langle 0| \chi^\dagger (-\tfrac{i}{2}\tensor{\mathbf{D}})^{2}\psi|
Q\overline{Q}({}^1S_0^{[1]})\rangle^{(1)} + \cdots,
\nn
\eqa
with all the involved NRQCD matrix elements available now. It is
anticipated that (\ref{NRQCD:amp:everything}) will faithfully
reproduce the infrared behavior of (\ref{QCD:ampls:NLO:alphas}).

\section{Matching the short-distance coefficients at NLO in $\bm{\alpha}_{\bm s}$}
\label{Matching:short-distance:NLO}

In section~\ref{tree:amplitude:matching} we have already conducted
the perturbative matching at tree level, and determined the
tree-level coefficients $c_i^{(0)}$ ($i=0,2$) in
(\ref{c0:and:c2:tree}). Substituting their explicit values into
(\ref{NRQCD:amp:everything}), and subtracting the contribution of
the tree-level NRQCD amplitude ${\mathbb A}_{\rm NRQCD}^{(0)}$, one
finds that the NRQCD amplitude at ${\cal O}(\alpha_s)$ reads:
\bqa
{\label{NRQCD-amp-NLO1}}
{\mathbb A}_{\rm NRQCD}^{(1)} &=& \sqrt{2 N_c}\Bigg\{ c^{(1)}_0 + c_2^{(1)} v^2
+ {4 C_F e_Q^2\alpha \alpha_s \over m}
\bigg\{{2v^2 \over 3}\left({1\over \epsilon_{\rm IR}}-\gamma_E +\ln 4\pi\right)
\nn\\
&+& {1\over 4v}\bigg(1+\frac{5}{6}v^2\bigg) \bigg[\pi^2+i\pi
\left(-{1 \over \epsilon_{\rm IR}}+\gamma_E +\ln{{\bf q}^2\over
\pi\mu^2} \right)\bigg] + \widetilde{C}_q v^2 \bigg\}\Bigg\}.
\eqa
Here $\widetilde{C}_q$ is a scheme-dependent imaginary constant. The
origin of this constant can be easily traced, which arises from the
last term in (\ref{NRQCD:amp:everything}), that is, the
$O(\epsilon)$ piece in the tree-level coefficient $c_2^{(0)}$
multiplied by the imaginary infrared pole in the matrix element
$\langle 0| \chi^\dagger (-\tfrac{i}{2}\tensor{\mathbf{D}})^{2}\psi|
Q\overline{Q}({}^1S_0^{[1]})\rangle^{(1)}$ in
(\ref{NRQCD:ME:v2:oneloop}).

Remarkably, $\widetilde{C}_q$ turns out to exactly equal the
scheme-dependent constant $C_q$ that appears in the NLO QCD
amplitude ${\mathscr A}^{(1)}_2$ in
(\ref{QCD:ampl:NLO:v:NLO:alphas}), which vanishes in $q_4$ scheme,
but equals ${i\pi \over 36v}$ in $q_D$ scheme. Clearly, the equality
of these two constants is by no means a coincidence.

Following (\ref{Matching:formula:ampl:QQbar:2gamma}) and
(\ref{full:amplitude:S:wave:series}), one readily recognizes  the
matching condition at order $\alpha_s$:
\bqa
{\mathscr A}_0^{(1)}+{\mathscr A}_2^{(1)}v^2 ={\mathbb A}^{(1)}_{\rm
NRQCD}.
\label{Matching:formula:NLO:alphas}
\eqa

Plugging the analytic expressions of ${\mathscr A}_i^{(1)}$
($i=0,2$), which are given in (\ref{QCD:ampls:NLO:alphas}), and the
expression of ${\mathbb A}^{(1)}_{\rm NRQCD}$, as given in
(\ref{NRQCD-amp-NLO1}), into this matching equation, one observes
that the infrared and Coulomb singularities, together with the terms
non-analytic in $|{\bf q}|$, exactly cancel from the both sides of
(\ref{Matching:formula:NLO:alphas}). This is of course as expected,
which is warranted by the general construction of the NRQCD
effective theory.

It is a nontrivial check for the consistency of our calculation,
that the imaginary parts, which receive the contribution solely from
the potential region, indeed fully cancel in both sides of
(\ref{Matching:formula:NLO:alphas}). In particular, it is remarkable
that the scheme-dependent imaginary pieces also cancel in both $q_4$
and $q_D$ schemes, thanks to the equality $\widetilde{C}_q=C_q$.
This cancelation is crucial for one to obtain the scheme-independent
short-distance coefficients~\footnote{We have made a further test of
the scheme-independence of the coefficient $c_2^{(1)}$. In the
QCD-side calculation, we attempt to use the alternative version of
the spin/color-singlet projector~\cite{Bodwin:2007ga}:
$$ \widehat{\Pi}_1^{(1)}(p,\bar{p})
= {1 \over 4\sqrt{2}E^2} (/\!\!\!{p}+m)\,\gamma_5 (/\!\!\!\bar{p}-m)
\otimes  {\mathbf{1}_c\over \sqrt{N_c}},$$
which is equivalent to (\ref{spin:singlet:projector}) only in $D=4$.
Following the same procedure as described in the text to extract the
${}^1S_0^{[1]}$ amplitude, we obtain identical results as in
(\ref{QCD:ampls:NLO:alphas}) except for the $C_q$ term. As expected,
one again gets $C_q=\widetilde{C}_q=0$ in $q_4$ scheme as before.
However, for the $q_D$ scheme in this case, one finds that
$C_q=\widetilde{C}_q= -{i\pi\over 18 v}$.}.

It is then straightforward to solve the order-$\alpha_s$
short-distance coefficients:
\begin{subequations}
\bqa
c_{0}^{(1)} &=&  {4\pi e_Q^2 \alpha \over m} {C_F\alpha_s\over\pi}\bigg({\pi^2\over 8}-{5\over 2}\bigg),
\label{c0:one:loop}
\\
c_2^{(1)}  &=&  {8\pi e_Q^2 \alpha \over 3 m} {C_F\alpha_s\over\pi} \bigg(
\ln {\mu^2\over 4m^2}
+ {19 \over 12}- {3\pi^2 \over 16}\bigg),
\label{c2:one:loop}
\eqa
\label{c0:and:c2:one:loop}
\end{subequations}
where $\mu$ is now interpreted as the factorization scale in the
$\overline{\rm MS}$ scheme. Note these coefficients are purely real.

With the knowledge of the $c^{(1)}_i$ in (\ref{c0:and:c2:one:loop}),
we can employ (\ref{Relation:F:G:c0:c2}) to determine the
short-distance coefficients $F$ and $G$ that appear in
(\ref{NRQCD:factorization:etaQ:decay:width}) through order
$\alpha_s$:
\begin{subequations}
\bqa
F({}^1S_0)& = & {m^2 \over 8\pi} \left[
|c_0^{(0)}|^2+ 2{\rm Re}\left(c_0^{(0)} c_0^{(1)*}\right)+
{\cal O}(\alpha_s^2)\right]
\nn\\
&= &  2\pi e_Q^4 \alpha^2\left[1+{C_F\alpha_s\over\pi}
\left({\pi^2\over 4}-5\right)+{\cal O}(\alpha_s^2)\right],
\\
G({}^1S_0)& = & {m^2\over 4\pi} {\rm Re}\left[c_0^{(0)} c_2^{(0)*}+
c_0^{(1)} c_2^{(0)*}+c_0^{(0)} c_2^{(1)*}+{\cal O}(\alpha_s^2)\right]
\nn\\
&=& -{8\pi e_Q^4\alpha^2 \over 3}\left[1+ {C_F\alpha_s \over \pi}
\left({5\pi^2\over 16}-{49\over 12}- \ln {\mu^2 \over 4m^2}\right)+{\cal O}(\alpha_s^2)\right].
\eqa
\label{Short:distance:coeffs:F:G:NLO:alphas}
\end{subequations}
The familiar result for $F({}^1S_0)$ through ${\cal O}(\alpha_s)$ is
then reproduced. The order-$\alpha_s$ correction to $G({}^1S_0)$ is
new, which constitutes the main result of this work.

To visualize the relative importance of these ${\cal O}(\alpha_s)$
corrections, one may choose $\mu=m$ in
(\ref{Short:distance:coeffs:F:G:NLO:alphas}):
\begin{subequations}
\bqa
F({}^1S_0) &=& 2\pi e_Q^4 \alpha^2 \left[1-3.38\times
{\alpha_s(m)\over \pi} + {\cal O}(\alpha_s^2)\right],\\
G({}^1S_0) &=& -{8\pi e_Q^4\alpha^2 \over 3}
\left[1+0.52\times\frac{\alpha_s(m)} {\pi}+{\cal
O}(\alpha_s^2)\right].
\label{G:1S0:NLO:alphas:numerical}
\eqa
\label{Short:distance:coeffs:F:G:Numerical}
\end{subequations}
The order-$\alpha_s$ correction to $F({}^1S_0)$ has modest effect
for $\eta_c$ decay to $\gamma\gamma$. If one takes
$\alpha_s(m_c)=0.3$, including this correction will reduce the
tree-level value of $F$ by about 30\%. Nevertheless, the
order-$\alpha_s$ correction to $G({}^1S_0)$ seems to have a much
minor impact, which only enhances the tree-level value of $G$
roughly by 5\% for $\mu=m$. Note even the relative sign of this
correction is uncertain. If one instead chooses $\mu=2m$, including
this ${\cal O}(\alpha_s)$ correction would {\it reduce} the
tree-level value of $G$ by about 10\%. It seems fair to state that,
at current level of experimental accuracy, the ${\cal O}(\alpha_s)$
correction to $G({}^1S_0)$ has very little phenomenological impact.

\section{Summary}
\label{summary}

NRQCD factorization approach provides a model-independent framework
for calculating the quarkonium annihilation decay rates. The
nonperturbative NRQCD matrix elements are universal, which can be
estimated from potential models or lattice QCD simulations, or
directly fitted from experiments. On the contrary, being
process-dependent, the short-distance coefficients appearing in the
NRQCD factorization formula can be systematically improved in
perturbation theory. To make a more precise prediction, it is
desirable to know the short-distance coefficients at higher accuracy
in $\alpha_s$ expansion.

In this work, we have computed the ${\cal O}(\alpha_s)$ corrections
to the short-distance coefficients relevant to the process of
pseudoscalar quarkonium decay to two photons. Specifically, we have
deduced the ${\cal O}(\alpha_s)$ piece of the coefficient
$G({}^1S_0)$, which is associated with the relative order-$v^2$
NRQCD matrix element in
(\ref{NRQCD:factorization:etaQ:decay:width}). This coefficient is
determined through equating the full QCD amplitude and the NRQCD
amplitude for the quark process
$Q\overline{Q}(^1S_0)\to\gamma\gamma$ through NLO in $\alpha_s$ and
$v^2$. As a consistency check, it is found that the IR and Coulomb
singularities, as well as those terms nonanalytic in $|{\bf q}|$,
are exactly canceled upon matching the calculations on both sides.
We have presented detailed descriptions for some subtle technical
issues encountered in the calculation, such as consistent
prescription for $\gamma_5$ in dimensional regularization, and the
ambiguity about the dimensionality of relative momentum when
projecting out the $S$-wave amplitude. We have explicitly verified
that different artificial schemes in extracting the $S$-wave
amplitude actually lead to the identical short-distance
coefficients.

Although the ${\cal O}(\alpha_s v^2)$ correction appears to be
phenomenologically insignificant for pseudoscalar quarkonium decay
to two photons, it is of interest to investigate the analogous
correction for the process of pseudoscalar quarkonium inclusive
decay to light hadrons.
 
{\noindent \bf Note added in the proof.} Shortly after this paper
was sumbitted, a similar work has appeared in arXiv, which also
addresses the ${\cal O}(\alpha_s v^2)$ correction to pseudoscalar
quarkonium decay to two photons~\cite{Guo:2011tz}. These authors
performed the matching calculation directly at the decay rate level,
and their final results agree with our
(\ref{Short:distance:coeffs:F:G:NLO:alphas}).

\begin{acknowledgments}
This research was supported in part by the National Natural Science
Foundation of China under grants No.~10875130, 10935012.
The research of W.~S. was also supported by Basic Science Research
Program through the NRF of Korea funded by the MEST under contract
No. 2011-0003023.
\end{acknowledgments}

\appendix

\section{Useful one-loop scalar integrals
\label{appendix}%
}

In this Appendix we tabulate those one-loop scalar integrals that
are encountered in the calculation of the NLO QCD correction to the
process $Q\overline{Q}({}^1S_0^{[1]})\to\gamma\gamma$. First we
adopt the following definitions of the 2-, 3-, 4-point
Passarino-Veltman scalar integrals~\cite{Ellis:2007qk}:
\begin{subequations}
\bqa
& & B_0(p_1^2;m_1^2,m_2^2)\equiv {\mu^{2\epsilon}\over i\pi^{D/2}} \Gamma(1-\epsilon)
\int\!\! d^D l\,{1\over (l^2-m_1^2)((l+q_1)^2-m_2^2)},
\\
& & C_0(p_1^2,p_2^2,p_3^2;m_1^2,m_2^2,m_3^2) \equiv {\mu^{2\epsilon}\over i\pi^{D/2}} \Gamma(1-\epsilon)
\nn\\
&& \qquad \times \int\!\! d^D l\,{1\over (l^2-m_1^2)((l+q_1)^2-m_2^2)((l+q_2)^2-m_3^2)},
\\
& & D_0(p_1^2,p_2^2,p_3^2,p_4^2; s_{12},s_{23}; m_1^2,m_2^2,m_3^2,m_4^2)
\equiv {\mu^{2\epsilon}\over i\pi^{D/2}} \Gamma(1-\epsilon)
\nn\\
&& \qquad \times \int\!\! d^D l \,{1\over (l^2-m_1^2)((l+q_1)^2-m_2^2)((l+q_2)^2-m_3^2)((l+q_3)^2-m_4^2)},
\eqa
\end{subequations}
where the spacetime dimension $D=4-2\epsilon$,
$q_n=\sum_{i=1}^n p_i$ and $q_0=0$ and $s_{ij}=(p_i+p_j)^2$.
The $i\epsilon$ prescription in each propagator has been tacitly assumed.

The desired results are:
\begin{subequations}
\bqa
&& B_0(-E^2 \pm 2k_1\cdot q+q^2; 0,m^2)= {1\over \epsilon_{\rm UV}}+\ln{\mu^2\over m^2}+2-\frac{2(E^2 \mp k_1\cdot
q)\,\ln {2(E^2 \mp k_1\cdot q)\over m^2}}{2(E^2 \mp k_1\cdot q)-m^2},
\\
&& C_0(0,m^2,-E^2+q^2 \pm 2k_1\cdot
q; m^2,m^2,0)= { {\rm Li}_2 \left({-E^2+q^2 \pm 2k_1\cdot
q\over
m^2}\right)-{\pi^2\over6}\over 2(E^2\mp k_1\cdot q)},
\\
&& C_0(4E^2,0,0;m^2,m^2,m^2)= {1\over 4E^2}
\Bigg({1\over2}\ln^2{1+ \beta\over 1-\beta}-{\pi^2\over2}-i\,\pi\ln{1+ \beta\over 1-\beta}\Bigg),
\\
& & C_0 (4E^2,m^2,m^2; m^2,m^2,0)= {1 \over 4E^2 \beta}
\Bigg\{\left({1\over \epsilon_{\rm IR}}+\ln{\mu^2\over m^2}\right)\left(-\ln {1+\beta\over 1-\beta}+i\,\pi\right)-
\pi^2
\nn\\
&& + {\rm Li}_2\left({2\beta \over 1+\beta}\right)
- {\rm Li}_2\left(-{2\beta \over 1-\beta}\right) -
i\,\pi\ln{4\beta^2\over 1-\beta^2} \Bigg\},
\label{infra:div:C0:integral}
\\
& & D_0(m^2,0,0,m^2; -E^2 +q^2\pm 2k_1\cdot q, 4E^2; 0,m^2,m^2,m^2)
={1\over 8E^2(E^2 \mp k_1\cdot
q) \beta}
\nn\\
&&\times \Bigg\{\left(\ln {1+\beta\over 1-\beta} -i\,\pi \right)
\bigg( {1\over \epsilon_{\rm IR}} +\ln{\mu^2\over m^2} - 2\ln{E^2 \mp k_1\cdot
q \over m^2}- \ln 4\beta^2 \bigg)+{\pi^2\over2}
\nn\\
&&+
2\left[ {\rm Li}_2\left({1-\beta \over 1+\beta}\right)-
{\rm Li}_2\left(-{1-\beta \over 1+\beta}\right)\right]\Bigg\}.
\label{infra:div:D0:integral}
\eqa
\end{subequations}
All the involved kinematic factors, such as $k_1$, $q$, $E$, $\beta$,
have already been defined in Section~ \ref{kinematics:definition}.
The dilogarithm is defined through
\bqa
{\rm Li}_2(x)=-\int_0^x \!\! dt \, {\ln(1-t) \over t}.
\eqa

All the infrared divergent one-loop scalar integrals (up to 4-point)
have been classified in Ref.~\cite{Ellis:2007qk}, which also
compiles the corresponding analytic expressions. We thus can
directly deduce the analytic expressions
(\ref{infra:div:C0:integral}) and (\ref{infra:div:D0:integral}) from
\cite{Ellis:2007qk} by making appropriate substitutions. The
analytic expression for the Coulomb-divergent three-point integral,
(\ref{infra:div:C0:integral}), is well known and can be found in
many places, {\it e.g.}, Eq.~(4.14) in \cite{Ellis:2007qk}, or
Eq.~(B10) in \cite{Bodwin:2008vp}. However, if one starts from
Eq.~(4.47) of \cite{Ellis:2007qk} to deduce the Coulomb-divergent
$D_0$ function, one would unfortunately end up with an erroneous
expression for the imaginary part. In (\ref{infra:div:D0:integral}),
we present the correct analytic expression for the imaginary part,
which is derived by employing the $D$-dimensional Cutkosky's rule.
Its correctness has been numerically verified with the help of the
\textsc{Mathematica} package \textsc{LoopTools}~\cite{Hahn:1998yk}.


\end{document}